\documentclass[pre,twocolumn,a4paper,showpacs]{revtex4-1}
\usepackage{amssymb}
\usepackage{wasysym}
\usepackage{graphicx}
\usepackage{epsfig}
\usepackage{subfigure}
\begin{document}

\title{Diffusion of innovations in  Axelrod's model}

\author{Paulo F.\ C.\ Tilles and Jos\'{e} F.\  Fontanari }

\affiliation{Instituto de F{\'\i}sica de S{\~a}o Carlos,
            Universidade de S{\~a}o Paulo,
            Caixa Postal 369, 13560-970 S\~ao Carlos SP, Brazil}

 
\begin{abstract}
Axelrod's  model for the dissemination of culture  contains two key factors required to model the process of diffusion of innovations, namely,  social influence (i.e., individuals become more similar when they interact) and homophily (i.e., individuals interact preferentially with similar others). The strength of these social influences are controlled by two  parameters: $F$, the number of features that characterizes the cultures and $q$, the common number of states each feature can assume. For fixed $F$, a large value of $q$ reduces the frequency of interactions between individuals because it makes their cultures  more diverse.
Here we assume  that the innovation is a new state of a cultural feature of a single individual -- the innovator -- and study  how the innovation spreads through  the networks among the individuals. For  infinite regular lattices in one (1D) and two  dimensions (2D), we find that initially  the successful innovation spreads  linearly with the time $t$, but in the long-time  limit it spreads diffusively ($\sim t^{1/2}$) in 1D
and sub-diffusively ($\sim t/\ln t$) in 2D. For finite lattices,  the growth curves for the number of adopters are typically concave
functions of   $t$. For random graphs with a finite number of nodes $N$, we argue that the classical S-shaped growth curves    result from a trade-off between  the average connectivity $K$ of
the graph and the per feature diversity $q$. A large $q$ is needed to reduce the pace of the initial spreading of the innovation and thus delimit the early-adopters stage, whereas a large $K$ is necessary to ensure the onset
 of the take-off stage at which the number of adopters grows superlinearly with $t$.  In an infinite random graph we find that
the number of adopters of a successful innovation  scales with $t^\gamma$  with  $\gamma =1$ for $K> 2$
and $1/2 < \gamma < 1$ for $K=2$.  We suggest that the  exponent $\gamma$ may be a useful index to characterize the process of diffusion of successful  innovations in diverse scenarios.
 
\end{abstract}

\maketitle

\section{Introduction} \label{sec:Intro}
An innovation is an idea, practice, or object that is perceived as new by a community.  In many situations, the diffusion of an innovation involves
one individual -- the  innovator -- who informs  potential adopters about a new idea and who, in turn,  pass on the
information to their near-peers. This exchange of information  occurs through a convergence  process in which the participants  share information with one another to reach a mutual understanding. The transfer of ideas occurs most frequently between two individuals who are alike  \cite{Rogers_62}. 

The above sketchy account of the process of diffusion of innovations, which is based on Everett Rogers'  1962 seminal book Diffusion of Innovations \cite{Rogers_62}, contains all ingredients of a very popular agent-based model proposed much later by the political scientist Robert Axelrod  \cite{Axelrod_97}  to explain the 
persistence of cultural diversity in the society  despite the effect of social influence that tends to increase the similarity of interacting agents.
 Since Axelrod uses the term culture  to indicate any  set of individual attributes that are susceptible to social influence, his model suits  well to study   the  diffusion  of innovations within   Rogers' framework.   
 
In Axelrod's   model   the agents  are represented by  strings  of  cultural features of length $F$, where each feature can adopt  $q$  distinct states (i.e., $q$ is  the common number of states each feature can assume).  A  characteristic that sets Axelrod's model  apart from most models of social influence (see \cite{Toral_07,Castellano_09,Galam_12} for reviews) is that it accounts  for homophily, which is  the tendency of individuals  to interact more frequently with individuals who  are more similar to them. In particular, according to the rules of Axelrod's model, the interaction between two neighboring agents occurs with probability proportional to the number of  cultural states they have in common: agents who do not have any cultural state  in common cannot interact and the interaction between agents who share all their cultural states does not result in any change. This is a most important characteristic from the perspective of the diffusion of innovations since the very nature of diffusion 
requires  some degree of heterophily  between the innovator  and the other  agents: ideally, they would be homophilous on all other variables (e.g., education and social status) even though they are heterophilous regarding the innovation \cite{Rogers_62}. In Axelrod's model an interaction  between two neighboring agents   consists of selecting at random one of the distinct features, and making the selected feature of  one of the agents -- the  target agent --  equal  to its neighbor's corresponding state. This rule models social influence  since the agents become more similar after they interact. Hence there is a positive feedback loop between homophily and social influence: similarity lead to interaction, and interaction leads to still more similarity \cite{Axelrod_97}.

In this paper we model the innovation as a novel state (say, state $q+1$) that appears in the initial configuration
 at a single cultural feature (say, feature $1$) of
a single agent (the innovator) located at the origin of a regular lattice of linear size $L$ with periodic boundary conditions or at the origin of a random graph of $N$ nodes. (In both cases the choice of the origin is arbitrary.)  In the regular lattice, $N=L^d$ where $d$ is the lattice
dimension. 
The $F$ cultural features of all other agents, as well as the $F-1$  features of the innovator, are set randomly according to a uniform distribution in the integers $1,2, \ldots, q$, as usual \cite{Axelrod_97}.
The important ingredient here is that  the innovation in the innovator (i.e., the state of feature 1 of the agent at the origin)  is fixed and cannot be changed by the rules of Axelrod's model. However, all other agents that adopt the innovation (adopters) may, in principle,  discard it through the interactions with their neighbors. This scheme is similar to the framework used to study cultural drift in Axelrod's model, modeled as a perturbation (noise) acting continuously at a single site of the lattice \cite{Klemm_03}, or to the presence of a zealot in the voter model  \cite{Mobilia_03}. The main difference is that both the noise source site and the zealot can change their states due to the interactions with their neighbors and their eventual return to the original states is guaranteed by external mechanisms.

In the thermodynamic limit $N \to \infty$, the ultimate fate of the innovation depends on the set of parameters $F$ and $q$, since the choice of these parameters determine  whether the  population reaches a consensus regime in which all agents (or at least a macroscopic number of agents) share the same culture or the
population is fragmented in an infinity of cultural domains of microscopic size.   
The nonequilibrium  phase transition separating these two stationary regimes has been object of intensive research by the statistical physics community \cite{Castellano_00,Vilone_02,JEDC_05,Vazquez_07,Barbosa_09,Peres_15}. Clearly,  in our scenario of diffusion of innovations
the consensus regime corresponds to the situation in which the innovation is adopted by the entire population
(i.e., the introduction of the innovation in
the community was successful): since the innovator has the innovation fixed there cannot be consensus unless all agents adopt the innovation. The multi-cultural regime corresponds to the situation
in which the innovation is confined to a finite region around the innovator.

We find that in the consensus regime, where the introduction of the innovation is successful, the initial spreading of the innovation
is ballistic in the one-dimensional regular  lattice and diffusive in the two-dimensional case. More pointedly, we show that, regardless of the
topology and connectivity of the network,  initially  the  number of adopters always increases linearly  with the time $t$, and  we derive  an analytical expression for its (constant) rate of increase  $v = v \left ( F, q \right )$  for  short times. In the
time asymptotic limit, we find that  Axelrod's model in the presence of the innovator behaves essentially as the voter model in presence of a zealot, and so the number of adopters grows with $t^{1/2}$ for
the one-dimensional lattice and with  $t/\ln t $ for the two-dimensional lattice \cite{Mobilia_03}.

However, in regular lattices the typical growth curves for the number of adopters   do not exhibit the classical S-shape observed in the  innovation
diffusion experiments  \cite{Rogers_62,Iglesias_12}. The sigmoid-shaped growth curves are recovered for 
random graphs of large average connectivity $K$. For random graphs with an infinite number of nodes we find that the number of adopters increases with $t^\gamma$ where $\gamma = 1$  for  $K>2$ irrespective of  the values of the parameters $F$ and $q$,
provided the convergence to the consensus regime is guaranteed. For $K=2$ we find $1/2  < \gamma < 1$. Since the  exponent $\gamma$ is sensitive to the social network topology it may be a useful quantitative measure to characterize the process of diffusion of innovations in
diverse scenarios.

The remainder of the paper is organized as follows. In section \ref{sec:model} we present a brief account of Axelrod's model and describe
how the model is modified  so as to reckon with the presence of the innovator at the origin of the network. In section \ref{sec:1d} we study the spreading of the innovation in the one-dimensional lattice and present the analytical calculation of the rate  $v$ for short times.  In addition, we   estimate the diffusion constant in the time asymptotic limit by assuming  that the Axelrod  model in the  presence of the innovator approaches the  consensus regime in a similar way as the voter model \cite{Mobilia_03,Evans_93,Frachebourg_96}. The spreading in the two-dimensional lattice is considered in section \ref{sec:2d} and the spreading in  random graphs in section \ref{sec:random}.
Finally, section \ref{sec:conc} offers our concluding remarks.

\section{Model}\label{sec:model}

In  Axelrod's model each agent is characterized by a set of $F$ cultural features and each feature   can take on $q$  different  states, which we label by the integers $1, 2, \ldots, q$. Hence there are $q^F$  distinct cultures in total.
In the initial configuration ($t=0$) each agent is assigned one of these cultures with equal probability. 
The $N$ agents are fixed in the nodes (sites) of a network. Here we will consider the one-dimensional ($d=1$) and the two-dimensional ($d=2$)  regular lattices of linear length $L$ with periodic boundary conditions,  i.e., a ring and a  torus, respectively,
where only nearest-neighbors interactions are allowed. To probe the effects of the short-ranged interactions and of the low average connectivity  of those regular lattices we will consider random graphs with average connectivity $K=2$ and $K=40$, as well.

According to the dynamics of the original model \cite{Axelrod_97},
at each time $t$ we pick an agent at random -- the target agent -- as well as one of its neighbors.  As usual in such asynchronous update scheme we choose the time unit as $\Delta t = 1/N$.  These two 
agents interact with probability equal to  their cultural similarity, defined as the fraction of 
common cultural features. As pointed out before, this  procedure models homophily.  An interaction consists of selecting at random one of the distinct features, and making the
selected feature of the target agent equal  to its neighbor's corresponding state. 
We note that neighboring agents with antagonistic cultures (i.e., cultures that do not share any cultural state)
 do not interact, whereas in the case the two agents are identical, the interaction produces no changes.  This procedure is repeated until 
the system is frozen into an absorbing configuration.  At an absorbing configuration  any pair of neighbors are either identical  or completely different  regarding their cultural states.

Our scenario to study the diffusion of innovations in Axelrod's model requires a simple modification of the original setup described above. In particular, we assume that there is a special agent -- the innovator -- located at the origin of the network who exhibits a novel cultural state in one of its cultural features. Without loss of generality, at the beginning of the simulation we assign the new state $q+1$ to  the cultural feature 1 of the innovator; the other $F-1$ features are set randomly to the usual $q$ states as done for the $F$ features of the other agents. Hence at $t=0$ the innovator and the rest of the population are heterophilous regarding the innovation. In addition, for the innovator only  the state $q+1$ (i.e., the innovation) is fixed and cannot be changed through the interaction with the other agents.  More specifically, in the case the innovator is selected as the target agent and its cultural feature 1 is selected to change, we do not  implement the change.
Otherwise, all other  interactions follow the usual rules  of the Axelrod model. We stress that this exception holds for the innovator only. The novel state can be replaced by the other states when the other agents  are updated. Of course,  only  feature 1 of the agents can assume the novel state $q+1$ as the dynamics unfolds; the remaining $F-1$ features take on the usual states  $1, \ldots, q$.

The main measure we will consider in this paper  is the  mean number of agents who have adopted the innovation (adopters) at a given time $t$, denoted by  $N \xi \left ( t \right) - 1$, where  $\xi \left ( t \right) $ is the mean fraction of agents that exhibit the innovation (i.e., state $q+1$ at cultural feature 1).  Note that the innovator does not count  as an adopter.  In the one-dimensional lattice the adopters   form a compact domain of length  $N \xi$ around the innovator.

\section{Spreading in the one-dimensional lattice}\label{sec:1d}

Our aim in this section  is to understand how the  parameters $F$ and $q$ influence the diffusion of the innovation in a ring of $N=L$ agents which are allowed to interact with their two nearest neighbors only.  It is well-known that in the absence of the innovator at the origin of the lattice the  model exhibits a phase transition in the space of parameters $\left ( F,q \right )$ 
that separates the regimes characterized  by  consensus   and  multicultural absorbing configurations  \cite{Vilone_02,JEDC_05}. Typically, the consensus regime occurs in the region of small $q$ and large $F$, whereas the multicultural regime occurs for large $q$ and small $F$.
In particular, for $F=3$ the innovator-free model exhibits a consensus regime for $q = 2$ and a multicultural regime for $q > 2$ in the thermodynamic limit \cite{Vilone_02}.  We do not expect that the presence of the innovator would in any way  alter  this transition. However, the  spreading of the innovation should depend strongly  on whether the parameters of the innovator-free model  are set so as to favor
consensus  or cultural diversity.  

\begin{figure}[!h]
\includegraphics[width=0.48\textwidth]{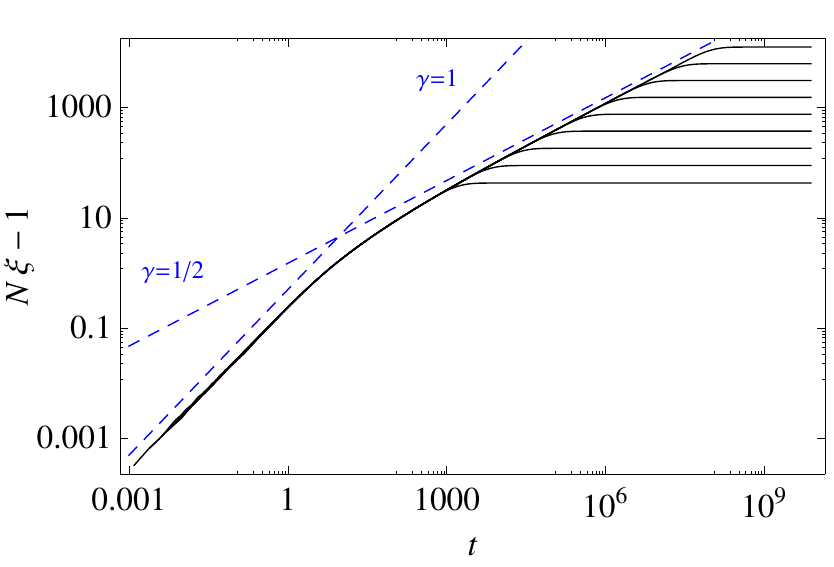}
\caption{(Color online) Mean number of adopters as function of  time   for chains of size 
(from bottom to top) $L = 2^{i} \times 50$, with $i = 0,1,\dots 8$ and parameters $F=3$ and $q=2$.    The dashed (blue) lines illustrate typical spreading behaviors in the one-dimensional lattice for which the number of adopters grows with $t$  (ballistic motion) and $t^{1/2}$ (diffusive motion). The total 
number of runs is $10^4$  for each $t$ and $L$.
 }
\label{fig:1}
\end{figure}

Figure \ref{fig:1} shows the mean number of agents that have adopted the innovation at time $t$  for the parameter set $F=3$ and $q=2$,
 which corresponds to the consensus phase of the innovator-free model   in the thermodynamic limit.  Of course, since for finite $L$ the number of adopters is bounded by the lattice size $N=L$, we must observe a saturation of the number of adopters in the asymptotic
limit $t \to \infty$. However, if we increase $L$ and keep $t$ finite (i.e., take the limit $L \to \infty$
 before the limit $t \to \infty$ \cite{Biral_15})  then all data fall on the  power law function $t^{\gamma}$ with $\gamma = 1/2$ for large $t$, which shows that in the asymptotic time limit the spreading of the innovation  is diffusive in an infinite lattice. 
 For small $t$, we observe a ballistic spreading of the innovation in the ring (i.e.,  $N \xi  - 1 \propto t^{\gamma} $ with $\gamma =1$) with a  rate $v$ that is independent of the lattice size. The challenge here is to derive an analytical estimate for the diffusion constant that characterizes the spreading for large $t$ and for the rate $v$  in the limit of small $t$. We will offer those estimates after we conclude the analysis of the  simulations.

\begin{figure}[!h]
\includegraphics[width=0.48\textwidth]{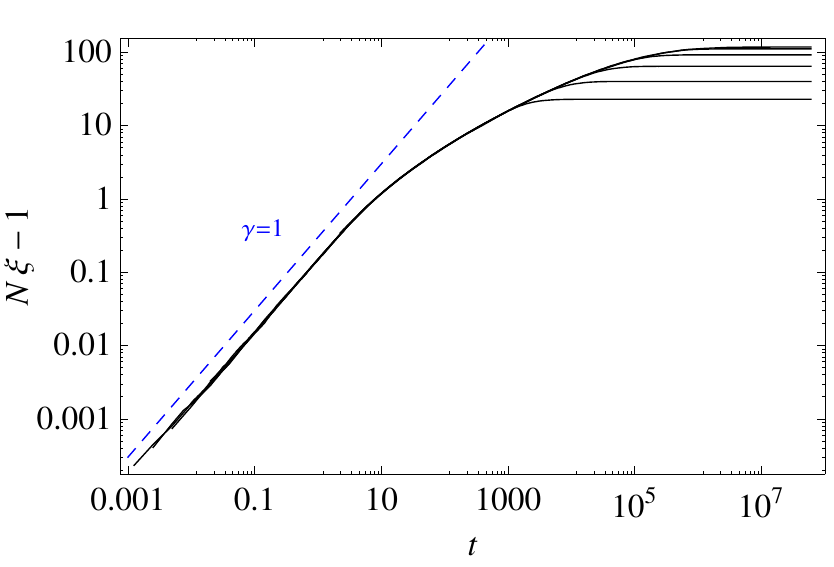}
\caption{(Color online) Mean number of adopters as function of  time  for chains of size 
(bottom to top) $L = 2^{i} \times 50$, with $i = 0,1,\dots 8$ and parameters $F=3$ and $q=3$. The curves are indistinguishable
for $L \geq 1600$ (i.e, $i  \geq 5$).   The dashed (blue) line illustrates the ballistic motion for which the number of adopters grows linearly with $t$. The total 
number of runs is $10^4$  for each $t$ and $L$.
 }
\label{fig:2}
\end{figure}

Figure \ref{fig:2} exhibits the evolution of the mean number of adopters for the parameter set   $F=3$ and $q=3$, corresponding to the multicultural regime of the innovator-free model in the thermodynamic limit.  Although for small $t$ we observe a ballistic spreading of the innovation,  the diffusive regime never sets in and the innovation is confined to only a  finite number of adopters even for an infinite lattice. In this case, the attempts of the innovator to introduce the  innovation in the community  failed altogether. 

\begin{figure}[!h]
\includegraphics[width=0.48\textwidth]{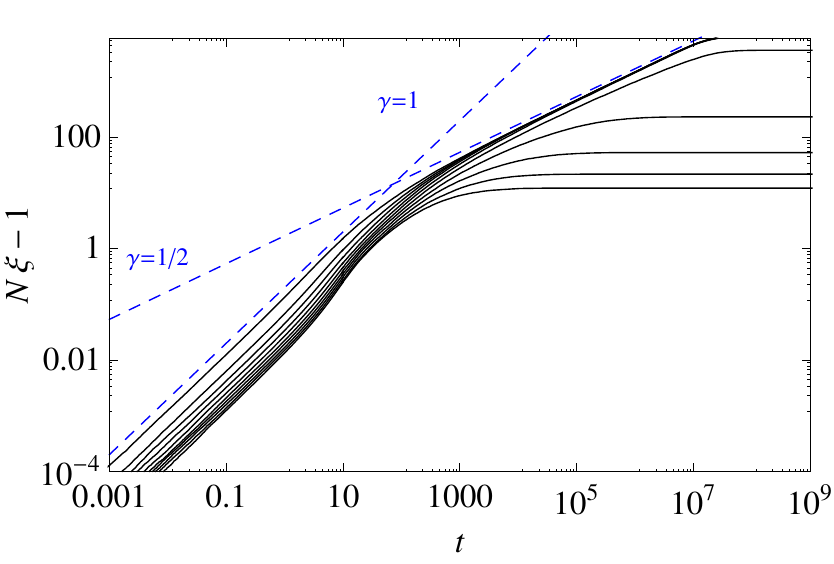}
\caption{(Color online) Mean number of adopters as function of  time  for a chain of size $L = N=25600$, $F=8$ and (bottom to top) $q=11,10, \ldots, 2$. The curves for $q=2,\ldots, 6$ are indistinguishable in the diffusive regime.
The phase transition occurs between $q=6$ and $q=7$.
 The dashed (blue) lines illustrate typical spreading behaviors in the one-dimensional lattice for which the number of adopters grows with $t$  (ballistic motion) and $t^{1/2}$ (diffusive motion). The total 
number of runs is $10^4$  for each $t$ and $q$.
 }
\label{fig:3}
\end{figure}

In order to punctuate the effect of varying the number of states $q$ for a fixed number of features $F=8$ and so observe the two regimes of
the innovator-free model in a same graph, Fig.\  \ref{fig:3} shows the evolution of the mean number of adopters for a chain of length $L=N=25600$. Since chains larger than this size produce indistinguishable curves, the figure exhibits effectively the results for
the thermodynamic limit in that range of $t$. The transition takes place between $q=6$ (consensus regime)  and $q=7$ (multicultural regime). This figure reveals  two  interesting results, namely, the diffusion constant does not depend on $q$ in the consensus regime and the rate of
increase  of the number of  adopters decreases with increasing $q$ for short times. Next we will offer analytical evidences for these findings.

\subsection{Analytical results for the short-time dynamics}

\begin{figure}[!h]
\includegraphics[width=0.48\textwidth]{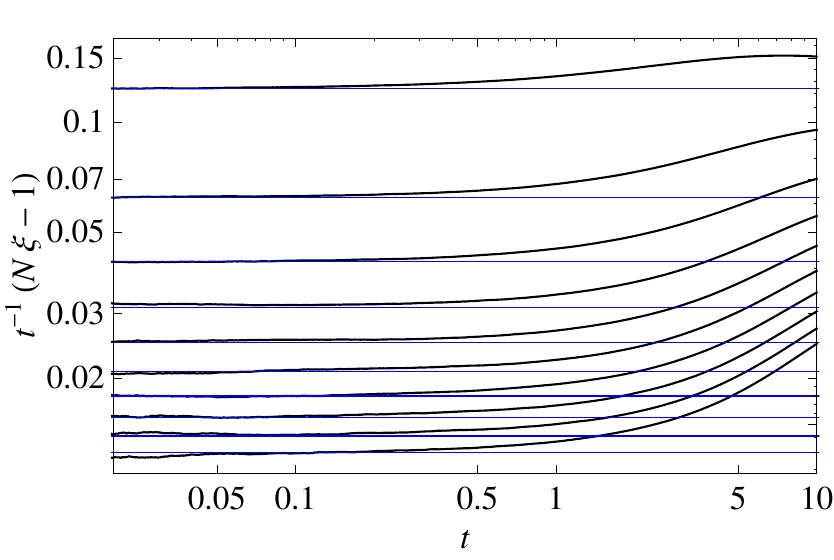}
\caption{(Color online) Short-time dynamics of the rescaled mean number of adopters  $\left ( N \xi - 1 \right )/t$ for a chain of size $L =25600$, $F=8$ and (top to bottom) $q=2,3, \ldots, 11$. 
 The horizontal (blue) lines show the predictions of eq.\ (\ref{v}). The total 
number of runs is $10^8$  for each $t$ and $q$.
 }
\label{fig:4n}
\end{figure}

The simulation results indicate that for small $t$ we can write $N \xi \left ( t \right ) \approx 1 + v t$ where
$v = v \left( F,q \right)$ is the short-time constant rate of increase of the number of adopters. We can determine $v$  by considering the first iteration only, i.e., the
evolution from $t=0$ to $t=1/N$. (We recall that    $ N \xi \left (  0 \right ) = 1$ and  $\Delta t = 1/N$).  The mean number of agents
exhibiting the innovation (i.e., the innovator plus the adopters) at $t=1/N$ can be written as
\begin{eqnarray}\label{n1}
N \xi \left ( 1/N \right )  & = & 2 \times \psi \left( F,q,N \right ) + 1 \times \left [ 1 - \psi \left( F,q,N \right )  \right ]  \nonumber \\
&  = & 1 + \psi \left( F,q,N \right )
\end{eqnarray}
where $\psi \left( F,q,N \right ) $ is the probability that the innovation is adopted by one of the neighbors of the innovator. Since at $t=0$ the
states of the $F$ features of all agents are set randomly according to a uniform distribution in the integers $1, \ldots, q$, except for feature
1 of the innovator that is set to $q+1$, we can easily write down this probability as
\begin{equation}\label{psi1}
\psi \left( F,q,N \right ) = \frac{1}{N} \sum_{m=1}^{F-1}   B\left ( F-1, 1/q \right )
 \frac{m}{F} \frac{1}{F-m} 
\end{equation}
where 
\begin{equation}
B\left ( F-1, 1/q \right ) = {{F-1} \choose m} \frac{1}{q^{m}} \left( 1- \frac{1}{q} \right)^{F-1-m}
\end{equation}
is a binomial distribution.
The factor $1/N$ is the probability of selecting any neighbor of the innovator as the target agent and then picking the innovator as the
interacting neighbor. Interestingly, this factor is not affected by  the connectivity of the agents. To see this let us  assume that the agents have
$K$ neighbors. Then the probability that we choose a target agent which is a neighbor of the innovator is $K/N$. The target agent  can interact with any of its
$K$ neighbors with equal probability and so the probability that it interacts with the innovator is $1/K$. The product of the probabilities of these two events then yields
the factor $1/N$ in eq.\ (\ref{psi1}).
Assuming the target agent shares $m=1,\ldots, F-1$ features  with the innovator then they will interact with probability $m/F$, but the target
agent will adopt the innovation only if feature 1 is the feature selected  to change among the $F-m$ different ones,   which happens
with probability $1/\left ( F- m \right)$. Finally, the binomial distribution weights the chance that the target agent will share exactly $m$ features with the innovator. The summation in eq.\ (\ref{psi1}) is carried out easily and yields
\begin{equation}\label{psi2}
\psi \left( F,q,N \right ) = \frac{1}{N} \frac{1 - q^{1-F}}{\left ( q-1 \right) F}.
\end{equation}
Since $t=1/N$ we have $N \xi \left ( 1/N \right ) = 1 + v/N$ from the definition of the rate $v$. Hence
\begin{equation}\label{v}
v  =  \frac{1 - q^{1-F}}{\left ( q-1 \right) F} .
\end{equation}
In Fig.\ \ref{fig:4n} we compare the predictions of eq.\ (\ref{v}) with the results of the Monte Carlo simulations for the short-time regime using  the same parameter set of Fig.\ \ref{fig:3}.  The analytical results agree perfectly with the simulation results, as expected.  The reason that $v$ decreases with increasing $q$ is because the probability of interaction decreases with $q$: the larger $q$ the lower the chance that the target agent and the innovator will have many features in common.  In addition, the larger $F$, the lower the chance
that the feature 1  of the innovator will be selected to be transferred to the target agent.
 Most importantly, as pointed out  above, eq.\  (\ref{v}) is valid for lattices of arbitrary dimension $d$ and for any graph topology since its derivation does not depend on the
connectivity of the agents. 

A word is in order about the short-time dynamics  of the standard voter model. This model  plays a key role in the
analysis of the diffusive regime as described next. In fact, since in the voter model the agents always interact and the interaction always
changes the target agent we have $\psi = 1/N$, i.e., the probability that the innovation is adopted by a neighbor of the innovator is given simply by the probability that any neighbor of the innovator is selected as the target agent and that  the innovator is the neighbor chosen to interact with the target agent. Hence $v=1$ for the voter model. 

Finally, we note that we have not proved that for short times the number of adopters increases linearly with $t$: what we have shown here was
that if we assume this linear growth, as observed in the Monte Carlo simulations,  then  the linear coefficient $v$  is given by eq.\ (\ref{v}).
However, as we will see next, for the one-dimensional voter model we can show analytically that the number of adopters in fact  grows linearly with $t$ and that $v=1$.

\subsection{Analytical results for the long-time dynamics}

An  unexpected and somewhat  unpleasant finding of our simulations is that the presence of the innovator at the origin increases considerably the time the dynamics takes to freeze into a consensus  absorbing configuration as compared with the freezing time in the innovator-free situation. More pointedly, we have observed that the agents first  reach a consensus for all features except for feature 1 (the innovation) and then they struggle to reach a consensus about  feature 1, which keeps fluctuating at the border of the domains
until the innovation eventually reaches fixation. This final stage of the dynamics  accounts for 
most of the  relaxation time. In this stage the dynamics reduces  essentially to the dynamics  of the voter model in which the agents are characterized by two states: $1$ if they adopt the innovation and $0$ otherwise. This is the reason the number of adopters in the  diffusive regime shown in Fig.\ \ref{fig:3} does not exhibit a dependence on $q$:  the system behaves as if   $q=2$ in that regime. There is, however, a  dependence on $F$ which appears because in the case of a single unfixed feature the probability of interaction  for Axelrod's model  is $\left (F-1 \right )/F$ whereas for the voter model it is $1$. But this fact can be accounted for by a simple rescaling of the time $t$.

To obtain a quantitative estimate of the diffusion constant  associated to the diffusive behavior exhibited in Figs.\ \ref{fig:1} and \ref{fig:3} we will calculate the diffusion constant $D_v$ for the voter model.  Let us assume that all agents in the chain exhibit opinion $0$ except for the innovator at the origin that exhibits opinion 1. The mapping of the voter model on a spin-$1/2$ ferromagnetic Ising chain with zero-temperature Glauber dynamics allows the derivation of the probability $p_n \left ( t \right )$ that an agent distant $n$ sites from the origin exhibits opinion 1 at time $t$
\begin{equation}\label{G}
p_{ n } \left( t \right) = 1 + e^{-t} \sum_{m=1}^{\infty} \left[ I_{\mid n \mid +m} \left(t \right) - I_{\mid n \mid -m} \left(t \right) \right],
\end{equation}
where $I_{n} \left(t \right)$ are the modified Bessel functions of the first kind  \cite{Glauber_63}. Using the property 
$ I_{-n} \left ( t \right ) = I_{n} \left ( t \right )$ 
we  verify that $p_0 \left ( t \right ) =1 $, as expected. Most importantly,  $p_n \left ( t \right )$ also  yields the probability that the  agents at the adjacent sites $k=1,2,\ldots,n-1$ exhibit the innovation too. In other words, if the agent at site $n$ has opinion 1  then all agents at the sites between the origin and site $n$ must have opinion 1 too.  Hence  $p_n \left ( t \right )$ is the probability that there are  at least $n$ adopters
to the right of the origin. The probability that there are exactly $n$ (contiguous) adopters to the right of the origin  at time $t$  is simply
\begin{equation}
q_{n} \left ( t \right ) = p_{n} \left ( t \right )  - p_{n+1}\left ( t \right )  =  e^{-t}  \left[ I_{n} \left(t \right) + I_{n +1} \left(t \right) \right],
\end {equation}
which is properly normalized since  $\sum_{n=0}^\infty q_{n}\left ( t \right ) =   p_0 \left ( t \right ) = 1$, as expected. Alternatively, we can verify
this normalization by  explicitly carrying out the summation over the indices of the Bessel functions  with the aid of the identity  \cite{Watson_22}
\begin{equation}\label{nor}
2 \sum_{n=1}^\infty  I_{n }\left ( t \right )  = e^{t} - I_{0} \left ( t \right ). 
\end {equation}
Since the spreading to the left of the origin is independent  of the spreading to the right  (the fixed origin acts as a wall that prevents information to pass  through it) we can say that the probability  that there are exactly $n$ (contiguous) adopters to the right of the origin and exactly $m$ to the left  at time $t$ is  $q_{n,m} \left ( t \right ) = q_n \left ( t \right ) q_m \left ( t \right )$. 
The mean number of adopters is then
\begin{eqnarray} \label{narr}
N \xi \left ( t \right ) -1   &  =  &  \sum_{n,m=0}^\infty (n+m) q_{n,m}  \left ( t \right ) \nonumber  \\
                             &  =  &  2 \sum_{n=0}^\infty n q_{n }\left ( t \right )  \nonumber  \\
                             &   = & 2  e^{-t}  \left[ 2 \sum_{n=1}^\infty n I_{n }\left ( t \right ) -  \sum_{n=1}^\infty I_{n }\left ( t \right ) \right ] .
\end {eqnarray}
Use of eq.\ (\ref{nor}) together with the identity   \cite{Watson_22}
\begin{equation}
2n I_{n} \left ( t \right ) = t \left [ I_{n-1} \left ( t \right ) - I_{n+1} \left ( t \right ) \right ]
\end {equation}
yields
\begin{equation}
N \xi \left ( t \right ) -1 = e^{-t} \left[ (2t+1) I_{0} \left( t \right) + 2t I_{1} \left( t \right)\right] -1,
\end {equation}
which is the exact expression of the number of adopters for the one dimensional voter model. Using the asymptotic form  of the Bessel functions for small arguments    we obtain
$N \xi \left ( t \right ) -1 \approx t$ in the short-time limit, which  proves that the initial  spreading is ballistic with rate $v=1$.
For large $t$ we can use the asymptotic form of the Bessel functions $I_n \left( t \right) \sim e^t/\sqrt{2 \pi t}$ to obtain
\begin{equation}\label{Nv}
N \xi \left ( t \right ) -1 \sim \left ( \frac{4 D_v}{\pi} \right )^{1/2} t^{1/2} 
\end {equation}
where $D_v = 2$ is the diffusion constant of the voter model. Since the number of adopters is akin to the total magnetization of the voter model, the diffusive behavior described by eq.\ (\ref{Nv})  concurs with the findings for the voter model  in the presence of a  zealot \cite{Mobilia_03}.

\begin{figure}[!h]
\includegraphics[width=0.48\textwidth]{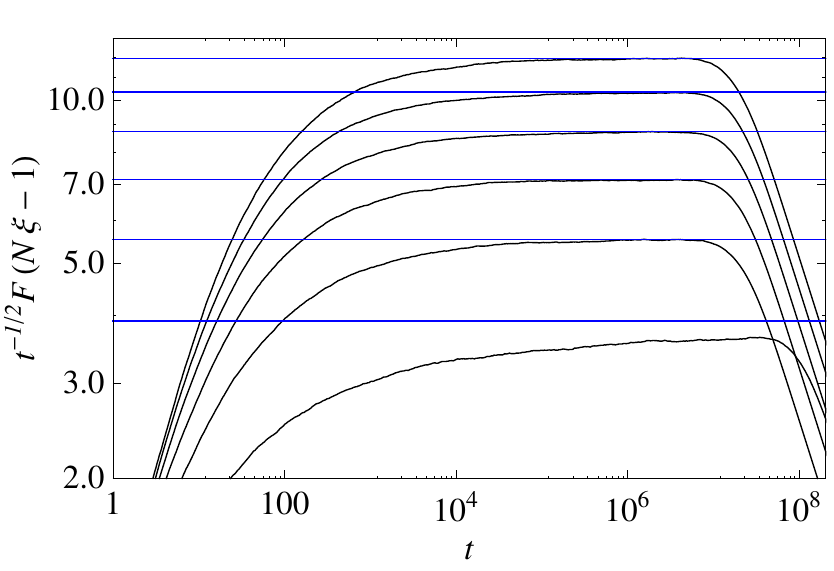}
\caption{(Color online) Long-time dynamics of the rescaled mean number of adopters  $F \left ( N \xi - 1 \right )/t^{1/2}$ for a chain of size $L = 6400$, $q=2$ and (bottom to top) $F=3,4,5,6,7, 8$.
 The horizontal (blue) lines show the predictions of eq.\ (\ref{Na}) with the diffusion constant $D = 2 \left ( F-1 \right)/F$. The total 
number of runs is $10^5$  for each $t$ and $F$.
 }
\label{fig:5n}
\end{figure}

As already pointed out, in the case of a single unfixed feature the probability of interaction  for Axelrod's model  is $\left (F-1 \right )/F$ whereas for the voter model it is $1$.  This means that  an interaction in Axelrod's model occurs at each $\left (F-1 \right )/F$ time steps on average. So the replacement $t \to t \left (F-1 \right )/F$ in eq.\ (\ref{Nv}) yields the correct expression for  the mean number of adopters in the diffusive regime of Axelrod's model,
\begin{equation}\label{Na}
N \xi \left ( t \right ) -1 \sim \left ( \frac{4D}{\pi} \right )^{1/2} t^{1/2} 
\end {equation}
where the diffusion constant is $D = 2 \left ( F-1 \right)/F$. 
Figure \ref{fig:5n} shows the comparison between the predictions of 
eq.\  (\ref{Na}) and   the results of the Monte Carlo simulations for a  chain of fixed size $L=6400$. We use this rather small chain  size because of the large number of runs
(typically $10^5$) needed to smooth out the fluctuations. However, except for $F=3$, this  figure shows a good agreement between theory and simulations. We note that the deviation from the diffusive regime  observed for very large values of $t$ is due to the saturation of the finite lattice. In the case $F=3$,
which corresponds to the lowest value of the  diffusion constant, much larger chain sizes are needed to observe the diffusive regime, which
we estimate would become perceptible for $t > 10^{12}$ only.
Hence Fig.\ \ref{fig:5n}  supports our conjecture that in the thermodynamic limit  the asymptotic spreading of the innovation in the one-dimensional Axelrod's model  is identical to the spreading in the voter model, except for a trivial rescaling of the  time or, equivalently, of the diffusion constant.

\section{Spreading in the two-dimensional lattice}\label{sec:2d}

\begin{figure}[!h]
\includegraphics[width=0.48\textwidth]{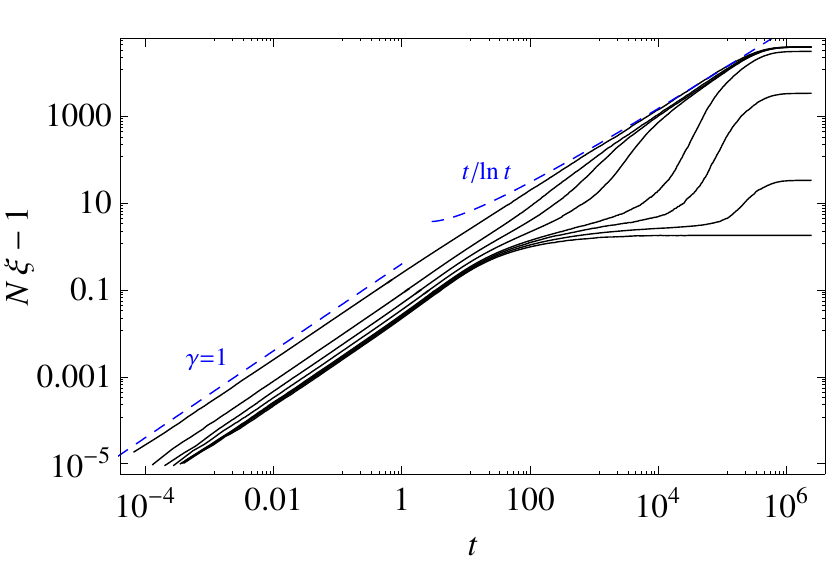}
\caption{(Color online) Mean number of adopters as function of  time  for a square lattice of linear size $L = 200$, $F=3$ and (bottom to top) $q=17,16, 15, 14, 11, 8, 5, 2$. For this lattice size, the curves for $q < 14$ are indistinguishable in the long-time regime. 
 The dashed (blue) lines illustrate the  linear  spreading ($\propto t$) of the innovation in the short-time regime and the slow 
 spreading  ($\propto t/\ln t$)  in the long-time regime, similar to the findings for the voter model in the presence of a zealot \cite{Mobilia_03}.
  The total 
number of runs is $10^4$  for each $t$ and $q$.}
\label{fig:4}
\end{figure}

We consider now the spreading of the innovation in a square lattice of linear size $L$. Figure \ref{fig:4} shows the results of the simulations for $F=3$ and a fixed  linear lattice size $L=200$. As in the one-dimensional case, the  presence of the innovator at the origin increases considerably the freezing time as compared with the innovator-free model, which poses then serious restrictions on the sizes of the lattices that we can simulate  and obtain  reliable statistical measures. For instance, for $F=3$ the innovator-free model   in the thermodynamic limit exhibits a consensus regime  for $q \leq 15$ and a multicultural regime for $ q \geq 16$ \cite{Barbosa_09}, whereas the results for $L = 200$ shown in the figure indicate that $q=15$ corresponds to the multicultural regime.  This misleading indication  is  due to  finite size effects.
Nevertheless, aside this discrepancy around the transition region, Fig.\ \ref{fig:4} reveals all the relevant information about the spreading of the innovation in the square lattice, as described next.

As pointed out, for small $t$ the mean number of adopters is given by $N \xi \left ( t \right ) -1 = vt$  with the rate $v$ given by eq.\ (\ref{v}), so that the innovation spreads faster  when the initial diversity of the population, which is  measured by $q$, is low. This dependence on $q < 14$ disappears for large $t$ indicating that, similarly to our findings for the one-dimensional lattice, the system behaves as a two-state voter model in the long time limit.  (In the thermodynamic limit this should happen for $q  <16$ \cite{Barbosa_09}.) Since in the context of the voter model the total number of adopters is analogous to the total magnetization of a lattice influenced by a zealot at the origin we conjecture that
$N \xi -1 \sim t/\ln t$  for large $t$ \cite{Mobilia_03}. This asymptotic behavior is consistent with the data presented in Fig.\ \ref{fig:4}, which
shows that the slopes  of the curves in the long-time regime (i.e., large $t$ but not too large to avoid the saturation  due to the finite value of $L$) are slightly less than 1.  In the multicultural regime $q > 15$ the innovation is restrained to a finite region around the innovator, as expected. A similar analysis  for  $F>3$ yields the same qualitative results.

\section{Spreading in random graphs}\label{sec:random}

A  conspicuous aspect of the diffusion of  innovations in a regular lattice discussed in the previous sections is the absence of the classical S-shaped growth curves in the plots of the number of adopters against time \cite{Rogers_62,Iglesias_12}. In fact, except perhaps for the region of values of  $q$ close to  the phase transition in the square lattice (see Fig.\ \ref{fig:4}) the growth curves do not exhibit the expected sigmoid shape. Of course, in the consensus regime this shape could be observed  for finite $N$  only, since  for an infinite lattice the number of  adopters does not saturate in the  limit $t \to \infty$. However, the manner $N\xi -1$ scales with $t$ in this limit offers a most valuable characterization of the spreading process, as shown in the previous sections for regular lattices.
To investigate whether the absence of the S-shaped growth curves  for the number  of adopters  in Axelrod's  model is due to the
short-ranged nature of the interactions  (nearest neighbors) or to  the low connectivity of the regular lattices, in this section we study 
the spreading of innovations in random graphs with $N$ nodes and  average connectivity  $K$.

Strictly speaking, the graphs we consider here are not the classical random graphs but the limit of sure rewiring of the Watts and Strogatz algorithm for constructing small-world networks \cite{Watts_98}. Explicitly, we begin  with  a  one-dimensional lattice of $N$ nodes, each node connected to $K$ neighbors ($K/2$ on each side), and periodic boundary conditions.  Then  for every node  $i= 1,\dots, N$  we rewire
the $K/2$  links between $i$ and $j=i+1,i+2,\ldots, i+ K/2$ (the sums are done modulo $N$) with probability $\beta =1$. Rewiring of a link is done by replacing the original neighbor of node $i$ by a random node chosen uniformly among  all possible nodes that avoid self-loops  and link duplication. The advantage of this formulation, which was also used in the studies of Axelrod's model on complex networks  \cite{Klemm_03b},
is that the origin (i.e., the node $i=1$, where the innovator is located) is guaranteed to be connected to at least  $K/2$ nodes, regardless of the values of $N$ and $K$, whereas for the classical random graphs that particular node has probability $e^{-K}$ of being  isolated from the
other $N-1$ nodes.
Otherwise, the resulting graphs are very similar to classical  random graph. For instance, the degree distribution is a Poisson distribution of mean $K$. We recall that, regardless of the topology and connectivity of the network, for small $t$ 
the mean number of adopters increases linearly with $t$   with a rate given by eq.\ (\ref{v}).  

\begin{figure}[!h]
\includegraphics[width=0.48\textwidth]{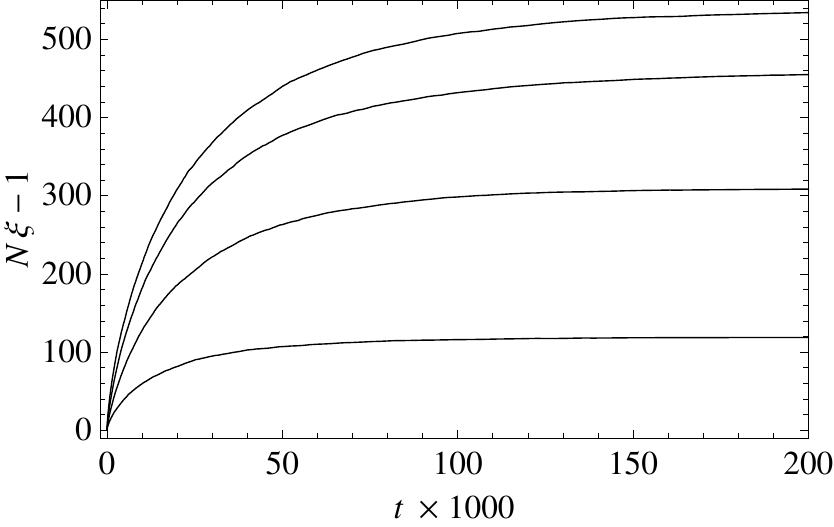}
\caption{(Color online) Mean number adopters as function of  time  for 
random graphs with $N=800$ nodes and  average connectivity $K=2$  for $F=3$ and (top to bottom) $q=2, 3, 4, 5$. The total 
number of independent runs is $10^4$  and in each run we used a different random graph.}
\label{fig:5}
\end{figure}

\begin{figure}[!h]
\includegraphics[width=0.48\textwidth]{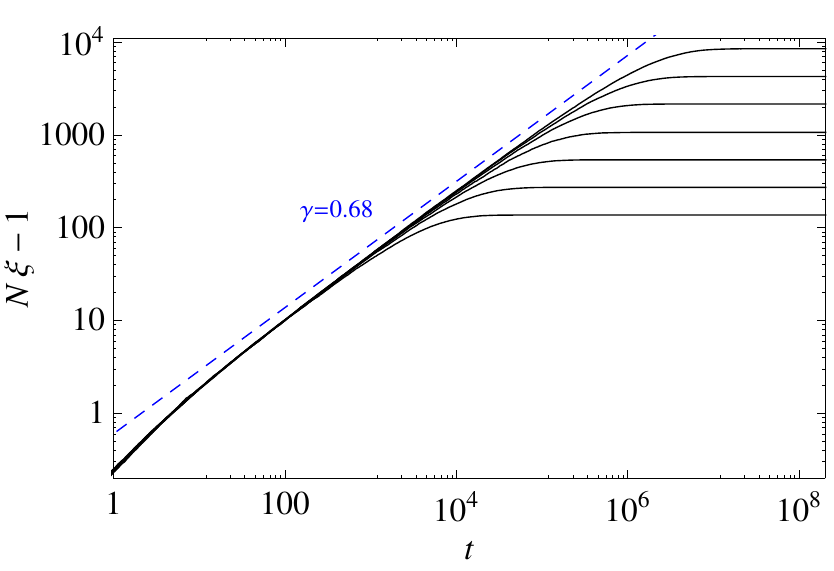}
\caption{(Color online) Mean number of adopters as function of  time  for 
random graphs with (bottom to top) $N= 200, 400, 800, 1600, 3200, 6400,12800$ nodes and  average connectivity $K=2$  for $F=3$ and  $q=2$. 
 The dashed (blue) line indicates that the number of adopters scales with $t^{0.68}$ in the limit $N \to \infty$. The total 
number of independent runs is $10^4$  and in each run we used a different random graph.}
\label{fig:6}
\end{figure}

\begin{figure}[!h]
\includegraphics[width=0.48\textwidth]{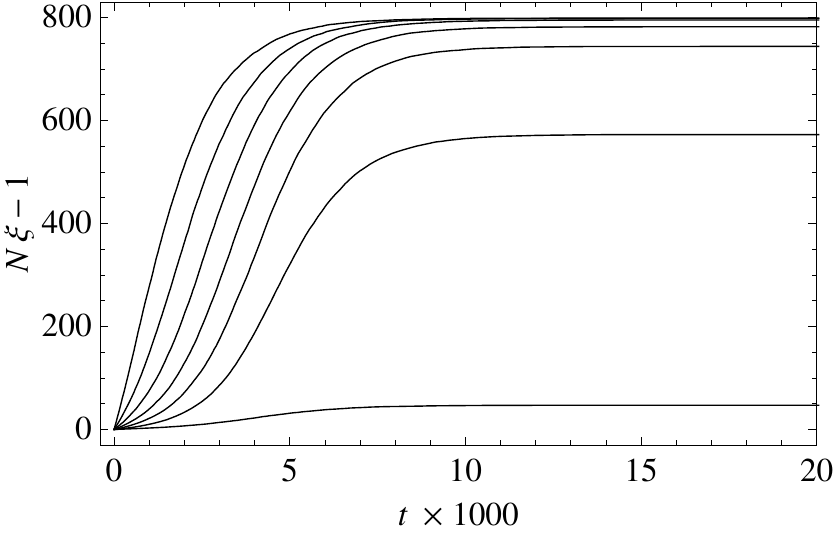}
\caption{(Color online) Mean number of adopters as function of  time  for 
random graphs with $N=800$ nodes and  average connectivity $K=40$  for $F=3$ and (top to bottom) $q=2, 4, 8, 16, 32, 64,128$. The total 
number of independent runs is $10^4$  and in each run we used a different random graph.}
\label{fig:7}
\end{figure}

\begin{figure}[!h]
\includegraphics[width=0.48\textwidth]{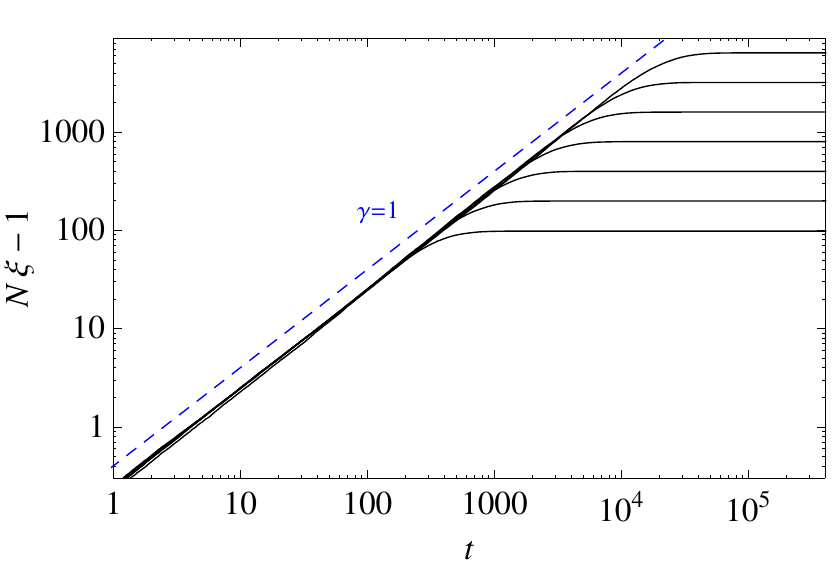}
\caption{(Color online) Mean number of adopters as function of  time   for 
 random graphs with (bottom to top) $N=100, 200, 400, 800, 1600, 3200, 6400$ nodes and  average connectivity $K=40$  for $F=3$ and  $q=2$. 
 The dashed (blue) line indicates that the number of adopters scales linearly  with $t$ in the limit $N \to \infty$. The total 
number of independent runs is $10^4$  and in each run we used a different random graph.}
\label{fig:8}
\end{figure}

\begin{figure}[!h]
\includegraphics[width=0.48\textwidth]{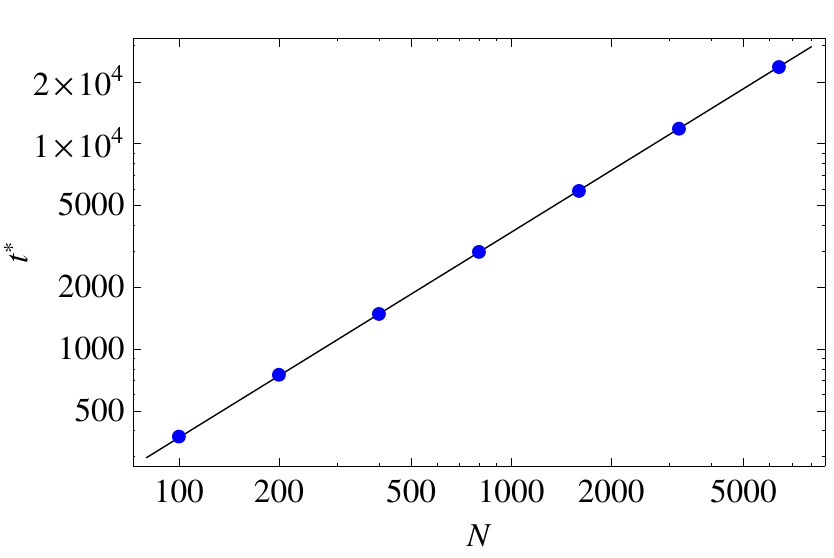}
\caption{(Color online) Mean relaxation time $t^*$ as function of the number of nodes $N$ of random graphs with average connectivity $K=40$  for $F=3$ and  $q=2$. The line is the fitting $t^* = 3.7 N$. 
 The total 
number of independent runs is $10^4$  and in each run we used a different random graph.}
\label{fig:9}
\end{figure}

In Fig.\  \ref{fig:5} we consider low  connectivity random graphs ($K=2$) for the purpose of comparison with the one-dimensional lattice. As in
that case, the growth curves do not exhibit inflection points. For such a low connectivity graph it is difficult to establish the existence of a  consensus regime because the graphs may not be connected,  so that the spreading of the innovation is restricted to the innovator's connected subgraph and, in this case, the asymptotic values of the number of adopters would be determined by the distribution of the sizes of the connected subgraphs. 
Nevertheless, since on the average the size of that subgraph increases with increasing $N$, it is still possible to observe the unlimited growth
of the number of adopters with increasing time in the thermodynamic limit. This is illustrated in 
 Fig.\  \ref{fig:6}  which shows that  the number of adopters  scales with $t^\gamma$ with $\gamma \approx 0.68$ for $F=3$ and $q=2$, indicating that the innovation spreads faster, due to the long range links, than in the chain with nearest neighbors interactions.  
For  $K=2$ we  find   the  same value of that exponent (i.e., $\gamma \approx 0.68 $)  for all $F > 2$ and $q$.  For  $F=2$ the consensus regime occurs for $q=2$ only, and in that case we  find $\gamma \approx 0.60$.
 
Figure \ref{fig:7} shows the adopters growth curves for high connectivity  random graphs ($K=40$) and $F=3$. This time the typical S-shaped curves show up provided the initial cultural diversity $q$ is not too small. We  did not find evidence of a multicultural regime  in the thermodynamic limit in this case. For instance, the large $t$ portion of the curve for  $q=128$  moves up as $N$ increases eventually  reaching the consensus regime  (see \cite{Klemm_03b} for a brief discussion of the phase transition in such random graphs). Figure \ref{fig:8} shows that in the consensus regime the number of innovators scales linearly with $t$. More generally and in contrast to the case $K=2$,  we find $\gamma = 1$ for $K > 2$ regardless of the parameters $F$ and $q$,  provided  that  the introduction of the innovation is
successful. It is also instructive to evaluate the mean time $t^*$ to reach a consensus absorbing configuration  for the data shown in Fig.\ \ref{fig:8}. In fact, Fig.\ \ref{fig:9} shows that the mean freezing time grows linearly with the number of nodes $N$, which is somewhat surprising since for a regular lattice the freezing time scales with the square of the number of sites \cite{Biral_15}. The high connectivity of
the graphs is probably accountable for this change in the scaling of $t^*$ with $N$. In addition, since only the consensus  regime is observed in our simulations we have $N \xi \to N$ in the very long-time (saturation) limit.

As in the analysis of the regular lattices, we find  that the long-time dynamics  in the consensus regime is not affected by the parameter
$q$. This is  expected:  in the initial configuration, $q$ yields  the number of different states that a given cultural feature can take on and as the dynamics proceeds towards a consensus absorbing configuration the social influence acts so as to decrease that number, leading eventually
to a situation, in the limit of very  large $t$, where each feature exhibits  only two distinct  states. Hence  in the consensus regime the asymptotic behavior  of the dynamics should not depend on $q$ or, more pointedly, for an infinite lattice   the model should behave as if $q=2$
in the limit $t \to \infty$.

\section{Conclusion}\label{sec:conc}

The focus of this study was on the spreading of an innovation in a decentralized diffusion system in which  the exchange of
information  among the potential adopters of an innovation is the main mechanism through which innovations spread \cite{Rogers_62}. Alternatively, we could also consider a centralized diffusion system  in which the innovation is presented to the individuals by an external source such as a global  media \cite{Shibanai_01}. In
the framework of Axelrod's model, the media is viewed as a global external field and its effects have been studied in great detail in
the statistical physics literature \cite{Avella_06,Candia_08,Avella_10,Peres_11,Peres_12}. 
In contrast to our findings for the decentralized diffusion system,
the relaxation to the absorbing configurations  in the presence of the external field is much faster than the relaxation in its absence, and  this very rapid relaxation results in concave growth curves for the number of agents who adopted the  states of the external field.

In fact, although Axelrod's model accommodates the two key ingredients -- homophily and social influence -- necessary to describe the spreading of an innovation in a community,   it does not always  reproduce the classical S-shaped   growth curves for the number of adopters of an  innovation \cite{Rogers_62}. For finite populations in general and  for infinite
populations in the multicultural regime,  the saturation part of the  sigmoid growth curves  is easily  reproduced in the limit of
large $t$, of course.  Since for short times the number of adopters grows linearly with $t$, regardless of the network topology,
the existence of an  inflection point requires an intermediate  stage  of superlinear growth. Such stage appears as a marginal quirk in the
one-dimensional lattice (see curve  for $q=11$ in Fig.\ \ref{fig:3}) which is imperceptible when plotted in a linear scale. In the square lattice, the superlinear growth stage appears only in a narrow  crossover region between the consensus and the multicultural regimes (see Fig.\ \ref{fig:4}). 

The requirement of an intermediate stage of superlinear growth for the existence of an inflection point in the adopters growth curves for finite systems  can be justified as follows. First, we note that for all the network topologies considered in this paper our simulations have revealed that the growth curves are linear in the short-time regime and that the linear coefficient $v$ is given by eq.\ (\ref{v}). Second, we note that 
for finite systems the number of adopters  saturates in the long-time regime, where the growth curves are  concave functions of
$t$. These two extremes can be joined continuously by a concave growth curve such as those shown in Fig.\ \ref{fig:5}, for which the first derivatives decrease continuously from the value $v$ at $t=0$ to the value $0$ in the limit $t \to \infty$. Third, we note that to exhibit an inflection point a growth curve must have  an intermediate region  where its first derivative is larger than $v$.  In fact, joining the short-time portion of the curve (derivative equal to $v$) with this intermediate portion (derivative greater than $v$) produces a region where the growth curve is convex.  Since close to the saturation regime the curve is concave there must exist an inflection point at which the concavity of
the curve changes. 
Finally, we note that since it is not possible to have a continuously differentiable growth curve by joining two linear segments (i.e., the segment with linear coefficient $v$ and a hypothetical  segment with linear coefficient greater than $v$), then the growth curve must exhibit a stage of superlinear growth.

Our analysis of the diffusion of innovations in random graphs  suggests that the desired S-shaped growth curves  emerge from the trade-off between the average connectivity $K$, which must be large enough to  guarantee a superlinear speedup of number of adopters, and the initial per feature  diversity $q$, which has the effect of  reducing the probability of interactions and hence of slowing down
the spreading of the innovation. The interplay between these parameters is illustrated in Figs.\ \ref{fig:5} and \ref{fig:7}. For  low connectivity random graphs (see Fig.\ \ref{fig:5}),
after the initial linear   spreading the number of adopters grows sublinearly with $t$ and so the growth curves are concave functions of 
the time regardless of the value of $q$.  This is probably an effect of the existence of many  weakly connected subgraphs with  bottlenecks that hinder the transmission of the innovation between them. 
The recipe to produce an inflection point for high connectivity random graphs is to begin the spreading with  a low  rate $v$ (the earlier
adopters stage), which can be achieved by  choosing a large value of  $q$, according to  eq.\ (\ref{v}), and wait till the number of adopters reaches a critical value (approximately 100 adopters  at $ t \sim 10^3$  for  the parameters of  Fig.\ \ref{fig:7}). The high connectivity of the graph will then ensure a superlinear spreading (the take-off stage)  and a  quick saturation (the later adopters stage) due to the finitude of the graph.

From a theoretical perspective, however, a more quantitative and unambiguous characterization of the process of diffusion of innovations  is obtained by  studying the approach to the consensus regime in an infinite lattice, i.e., by determining how the number of adopters
scales with $t$ in the limit $t \to \infty$. In particular, our Monte Carlo simulations indicate that in this limit Axelrod's model in the presence of the innovator behaves as the voter model in presence of a zealot, and so the number of adopters grows with $t^{1/2}$ for
the one-dimensional lattice and with  $t/\ln t $ for the two-dimensional lattice \cite{Mobilia_03}. In that sense, the spreading is diffusive in one dimension and sub-diffusive in two dimensions.
 For random graphs  with average connectivity $K>2$ we find that  the number of adopters grows with  $t^\gamma$ where $\gamma=1$. Our finding that $1/2 < \gamma < 1$ for $K=2$ indicates that the value of this exponent  is sensitive to the network topology and so it may be a useful  measure to compare the efficiency of the process of diffusion of successful innovations in different scenarios.

\acknowledgments
 The work of J.F.F. was partially supported  
 by  grant    2013/17131-0, S\~ao Paulo Research Foundation (FAPESP)
and by grant  303979/2013-5, Conselho Nacional de Desenvolvimento Cient\'{\i}fico e Tecnol\'ogico (CNPq).
P.F.C.T. was supported by grant  2011/11386-1, S\~ao Paulo Research Foundation (FAPESP).

\end{document}